# Complete Interband Transitions for Non-Hermitian Spin-Orbit-Coupled Cold-Atom Systems

Dong Liu[1], Zejian Ren[2,3], Wai Chun Wong[1], Entong Zhao[1], Chengdong He[1], Ka Kwan Pak[1], Gyu-Boong Jo[1]* and Jensen Li[1]*

[1]Department of Physics, Hong Kong University of Science and Technology, Clear Water Bay, Hong Kong, China

[2]Microelectronics Thrust, Hong Kong University of Science and Technology (Guangzhou), Guangzhou, China

[3]GuangZhou Nansha IT Park Postdoctoral Programme, Guangzhou, China

*Corresponding author. Email: gbjo@ust.hk and jensenli@ust.hk

**Abstract**

Recently, synthetic spin–orbit coupling has been introduced into cold-atom systems for more flexible control of the Hamiltonian, which was further made time-varying through two-photon detuning to achieve dynamic control of the cold-atom state. While an intraband transition can be adiabatically obtained, a complete interband transition, rather than a superposition of different bands, obtained through fast sweeping is usually guaranteed by having the positions of the initial and final states be far away from any band gap in the quasimomentum space. Here, by introducing an additional non-Hermitian parameter through an atom-loss contrast together with two-photon detuning as two controllable external parameters, both intraband and complete interband transitions can be achieved independent of the positions of the initial and final states. In addition, a "point-source diagram" approach in the 2D external parameter space is developed to visualize and predict the locations of any nonadiabatic transitions. This control protocol can have potential applications in quantum state control and quantum simulations using cold-atom systems.

# 1. Introduction

Recently, ultracold atoms in optical lattices have become an emerging platform for exploring various topological physics that can be challenging to achieve using conventional approaches [1, 2]. For instance, spin–orbit coupling (SOC), which is crucial in numerous condensed matter phenomena, is often predetermined in a given solid-state material but can now be engineered on demand in cold-atom systems using laser fields [3–7]. Moreover, atom loss can be purposely introduced into cold-atom systems [8–10] to study a wide range of nontrivial behaviors in non-Hermitian physics [11–29]. For instance, non-Hermitian phase transitions, exceptional points (EPs), chiral state transfer around EPs, and nonadiabatic transitions (NATs) [8–10] have now been demonstrated. As such, the Hamiltonian of a cold-atom system can now be flexibly controlled in both reconfigurable and dynamic manners, enabling precise engineering of the cold-atom state under a time-varying Hamiltonian.

Control of a quantum state through a time-varying Hamiltonian has a long history, particularly in the adiabatic regime [30]. When an external parameter of the Hamiltonian is changed sufficiently slowly, the quantum system adapts its state to the changing conditions and remains in the instantaneous eigenstate of the Hamiltonian. Such adiabatic control can be applied to a wide range of quantum systems [31–36]. In contrast, to trigger an interband transition, one can resort to Landau–Zener (LZ) tunneling [37–42] by rapidly passing through a band gap, e.g., in a two-band system. However, a complete interband transition occurs only when the positions of the initial and final states are both far from the band gap, so they are well approximated by the diabatic states [43]. Without this assumption, the final state is generally a superposition of the states of the two bands without a complete interband transition.

For cold atoms in an optical lattice, the lattice can be moved with a constant velocity or accelerated by changing external electromagnetic fields, e.g., by applying a time-varying frequency offset between two optical beams, equivalent to accelerating atoms in the lattice rest frame [37–42]. Inspired by previous works using atom loss as an additional external parameter [44–46], such a parameter is used to extend the Hamiltonian to the non-Hermitian regime as an interim process in the control. This approach provides more degrees of freedom for controlling the quantum state. An example is the protocol for a non-Hermitian shortcut to adiabaticity, which allows the system to perform a seemingly adiabatic intraband transition in a much shorter duration than the conventional adiabatic approach [47–49]. In this work, we will exploit such a non-Hermitian parameter (atom loss) for cold-atom state control. For simplicity, we focus on a two-band system in the presence of synthetic SOC. In addition to the two-photon detuning to change the quantum state of the original Hermitian system (effectively in the quasimomentum space), the atom loss will be turned on and off as an interim process to achieve complete interband transitions with flexible choice of the positions of the initial and final states in the quasimomentum space. Furthermore, we develop a point-source diagram approach to visualize and predict the locations of the transitions in the 2D external parameter space in the interim process. Such an extended non-Hermitian path for controlling a Hermitian system has potential applications in quantum state control and quantum simulations using cold-atom systems [8–10].

## 2. Cold-Atom System with Complex-Energy Bands

Our discussion starts from a cold-atom system with SOC [3–7, 10], as illustrated in Fig. 1 $(a)$. Two atomic hyperfine states of ultracold atoms in an optical trap are separated by the application of an external magnetic field, referred to as spin-up (red dots) and spin-down (blue dots) atoms. The system dynamics follow the Schrödinger equation $i\hbar \frac{d}{dt}|\psi\rangle = \mathcal{H}|\psi\rangle$, where the effective $2 \times 2$ Hamiltonian, in terms of the spin-up and spin-down components $\psi_\uparrow$ and $\psi_\downarrow$ of $|\psi\rangle$, can be expressed as

$$\mathcal{H} = \begin{pmatrix} \frac{\hbar^2}{2m}(q_x - k_r)^2 + \delta/2 - i\gamma_\uparrow/2 & \Omega_R/2 \\ \Omega_R/2 & \frac{\hbar^2}{2m}(q_x + k_r)^2 - \delta/2 - i\gamma_\downarrow/2 \end{pmatrix}. \quad (1)$$

Here, the cold atoms in the optical trap have quasimomentum $q_x$ in the $x$-direction, while two Raman beams contribute additional momentum $k_r = \frac{2\pi}{\lambda_{556}}\sin(\alpha/2)$ (intersecting at angle $\alpha$ with wavelength $\lambda_{556} = 556\ nm$, as shown by the green arrows in Fig. 1 $(a)$), giving rise to the real momentum $q_x \mp k_r$ for the two spins in the diagonal terms of $\mathcal{H}$. The two beams are actually detuned by $\pm\delta/2$ from the Raman resonance, contributing to opposite shifts in the diagonal terms of $\mathcal{H}$. The same pair of Raman beams also couples the two spin states through an intermediate energy level, introducing synthetic SOC with constant strength $\Omega_R/2$ as the off-diagonal terms of $\mathcal{H}$. In the Hamiltonian, the atom loss term $-i\gamma_\uparrow$ or $-i\gamma_\downarrow$ can be facilitated by another single near-resonant beam ("lossy" beam), which induces transitions from either the spin-up or spin-down state to other energy levels (not shown). Typically, the lossy beam interacts with two spins at different strengths, and in the following discussion, we assume that $\gamma_\downarrow > \gamma_\uparrow$. The terms with a square dependence on the momentum are related to energy through the factor $\hbar^2/(2m)$, where $m$ is the mass of an atom and $\hbar$ is the reduced Planck constant.

Here, we define the recoil momentum $p_r = \hbar k_r$ and the recoil energy $E_r = \hbar^2 k_r^2/2m$ as the natural choices of units. We further shift the energy by subtracting $E_0 = \hbar^2(q_x^2 + k_r^2)/2m - i(\gamma_\downarrow + \gamma_\uparrow)/4$ from both diagonal values of $\mathcal{H}$ to facilitate a discussion of the band structure. The shift in the real energy is chosen to align the mid gap to zero energy, while the shift in the imaginary part is chosen so that the system can be regarded as effectively Hermitian (still giving orthogonal eigenstates) when there is no loss contrast between the two spins ($\gamma_\downarrow = \gamma_\uparrow$). Then, the Schrodinger equation (in normalized time $\tilde{t} = t\Omega_R/(2\hbar)$ and Hamiltonian $\widetilde{\mathcal{H}}(\tilde{q}_x, \Delta\tilde{\gamma}) = 2\mathcal{H}/\Omega_R$) for the gauged wavefunction $|\tilde{\psi}\rangle = \exp(iE_0 t/\hbar)|\psi\rangle$ [17] is rewritten as

$$i\frac{d}{d\tilde{t}}|\tilde{\psi}\rangle = \widetilde{\mathcal{H}}(\tilde{q}_x, \Delta\tilde{\gamma})|\tilde{\psi}\rangle, \qquad (2)$$

with the effective Hamiltonian simplified as

$$\widetilde{\mathcal{H}}(\tilde{q}_x, \Delta\tilde{\gamma}) = \begin{pmatrix} (-\tilde{q}_x + i\Delta\tilde{\gamma})\frac{2E_r}{\Omega_R} & 1 \\ 1 & (\tilde{q}_x - i\Delta\tilde{\gamma})\frac{2E_r}{\Omega_R} \end{pmatrix}, \qquad (3)$$

where $\tilde{q}_x = 2q_x/k_r - \delta/(2E_r)$ and $\Delta\tilde{\gamma} = (\gamma_\downarrow - \gamma_\uparrow)/(4E_r)$ are defined as the two external parameters that we can tune, which are related to the quasimomentum and loss contrast between the two spins, respectively. Here, $\tilde{q}_x$ is actually tuned by two-photon detuning $\delta$, and we call this the quasimomentum parameter for ease of discussion. Since the self-adjointness of $\widetilde{\mathcal{H}}$ depends on whether $\Delta\tilde{\gamma}$ equals zero, we call $\Delta\tilde{\gamma}$ the non-Hermitian parameter. Without loss of generality, we choose $\Omega_R = 2E_r$ as an example. When there is zero loss contrast ($\Delta\tilde{\gamma} = 0$), two energy bands ($E_\pm$, two eigenvalues of $\tilde{\mathcal{H}}$) in $\text{Re}(E)$ form a band gap against $\tilde{q}_x$ due to the SOC, as shown on the vertical semitransparent plane in Fig. 1 (b). Besides, as $\tilde{\mathcal{H}}$ is self-adjoint, $\text{Im}(E) = 0$, as shown on the vertical semitransparent plane in Fig. 1 (c). When there is a loss contrast ($\Delta\tilde{\gamma} > 0$), each energy band becomes a surface in the 2D external parameter space ($\tilde{q}_x, \Delta\tilde{\gamma}$). As $\tilde{\mathcal{H}}$ is not self-adjoint, the two bands ($E_\pm$, with $\text{Re}(E_+) > \text{Re}(E_-)$) have both $\text{Re}(E)$ and generally nonvanishing $\text{Im}(E)$, where $\text{Re}(E)$ and $-\text{Im}(E)$ represent the energy and damping rate of the corresponding eigenstates. Here, we use the term "complex-energy bands" to denote the bands when $\tilde{\mathcal{H}}$ is not self-adjoint. When $\Delta\tilde{\gamma} = 1$ ($\gamma_\downarrow - \gamma_\uparrow = 2\Omega_R$) and $\tilde{q}_x = 0$, the two complex-energy bands are degenerate at the EP (purple dot), where both complex-energy bands and their corresponding eigenstates are degenerate, as shown in Fig. 1 (b, c). After finding the eigenstates at a fixed pair of external parameters ($\tilde{q}_x, \Delta\tilde{\gamma}$), we color the bands according to their spin polarization:

$$\langle S \rangle = \frac{|\psi_\uparrow|^2 - |\psi_\downarrow|^2}{|\psi_\downarrow|^2 + |\psi_\uparrow|^2}$$

which ranges from $-1$ (solely spin-down) to $+1$ (solely spin-up). The red complex-energy band ($\langle S \rangle > 0$) is the lower-loss band (with a larger $\text{Im}(E)$) because the "lossy" beam has a weaker interaction with the spin-up atoms, $\gamma_\uparrow < \gamma_\downarrow$.

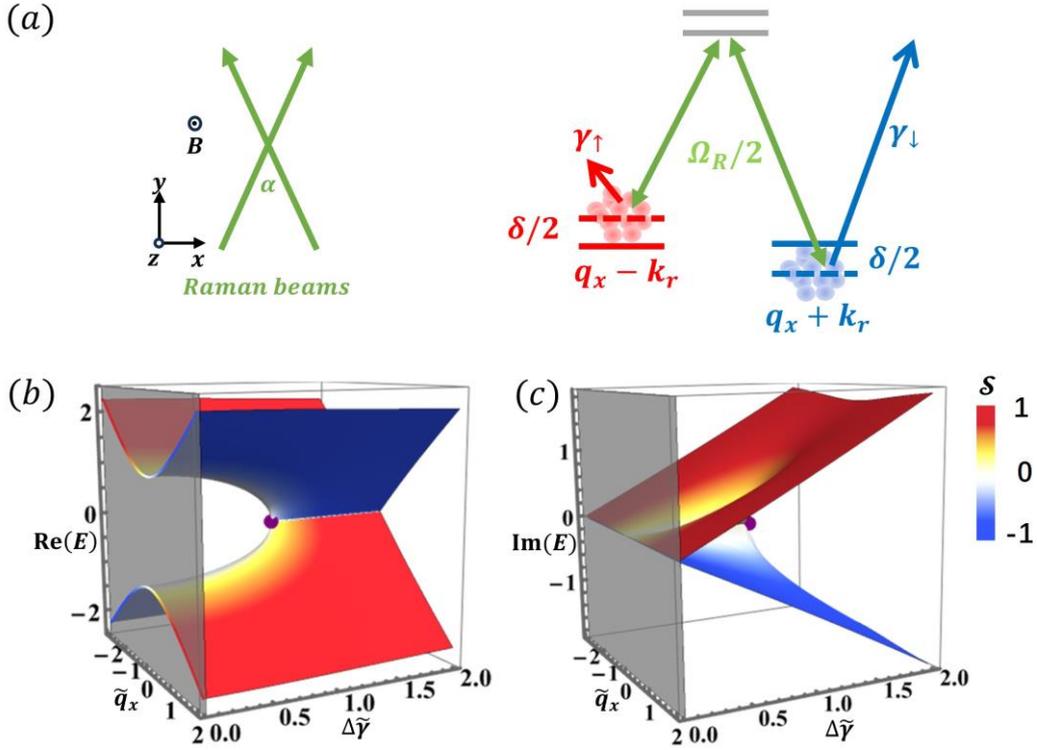

**Fig. 1** Panel $(a)$ shows a schematic diagram of the cold-atom system. SOC is induced by Raman beams (green arrows), and atom loss is induced by a single resonant beam. Panels $(b)$ and $(c)$ show $\text{Re}(E)$ and $\text{Im}(E)$ in the 2D external parameter space, respectively. On the semitransparent plane, two energy bands form a band gap at $\tilde{q}_x = 0$. With loss contrast $\Delta\tilde{\gamma} > 0$, the complex-energy bands are degenerate at the EP (purple dot), where $\Delta\tilde{\gamma}$ is equal to the coupling strength of $\widetilde{\mathcal{H}}$. The color of the bands indicates the spin polarization of the corresponding eigenstates.

## 3. Time-Varying Cold-Atom System

In this section, we aim to find methods for achieving complete interband transitions between two energy bands (on the semitransparent plane in Fig. 1 $(b)$) by dynamically changing the external parameters so that complete interband transitions can occur with flexible choice of the initial and final state positions in the quasimomentum parameter space. Specifically, we focus on two different cases of changing the external parameters. In the Hermitian case, with $\Delta\tilde{\gamma} = 0$, we change the quantum state of the Hermitian system by gradually changing the quasimomentum parameter $\tilde{q}_x$, following the black path in Fig. 2 $(a)$. In the non-Hermitian case, compared with the Hermitian case, the non-Hermitian parameter $\Delta\tilde{\gamma}$ will be induced to extend the original Hermitian system to the non-Hermitian regime, which serves as an interim process to assist interband transitions between two energy bands. For simplicity, we adopt the green path in Fig. 2 $(a)$, which consists of three straight lines, as the extended non-Hermitian path to demonstrate the complete interband transitions.

For a given path of changing external parameters in the 2D external parameter space, abbreviated as $\boldsymbol{q} = (\tilde{q}_x, \Delta\tilde{\gamma})$, we define $\boldsymbol{v} = d\boldsymbol{q}/d\tilde{t}$ to denote the time rate change of the external parameters, which is externally specified at every point on the path as a "velocity" vector: $\boldsymbol{v} = (v_q, v_\gamma) = \left(\frac{d}{d\tilde{t}}\tilde{q}_x, \frac{d}{d\tilde{t}}\Delta\tilde{\gamma}\right)$. We note that our external parameter space is actually not the phase space in the conventional meaning and that $\boldsymbol{v}$ is not the phase-space velocity field (Wigner current velocity) as a function in the quantum phase space [50–53] for representing the dynamics of the state. By substituting $\frac{d}{d\tilde{t}} \to iv_q \frac{\partial}{\partial \tilde{q}_x} + iv_\gamma \frac{\partial}{\partial \Delta\tilde{\gamma}}$ into Eq. (2),

$$iv_q \frac{\partial}{\partial \tilde{q}_x}|\tilde{\psi}\rangle + iv_\gamma \frac{\partial}{\partial \Delta\tilde{\gamma}}|\tilde{\psi}\rangle = \widetilde{\mathcal{H}}(\tilde{q}_x, \Delta\tilde{\gamma})|\tilde{\psi}\rangle,$$

we can see that the externally specified $\boldsymbol{v}$ for a given path determines the dynamics of the state through the equation of motion.

In both the Hermitian and non-Hermitian cases that we will discuss (black and green paths in Fig. 2 $(a)$), we adopt a constant "velocity" $|\boldsymbol{v}| = \sqrt{v_q^2 + v_\gamma^2}$ to change the two external parameters, and the total time spent on the whole path is determined through $T = \int |d\boldsymbol{q}|/|\boldsymbol{v}|$. We would like to evaluate how similar the final state is to the state of the target energy band by using the inner product

$$\langle E(T) \rangle = \frac{\langle \tilde{\psi}(T)|\widetilde{\mathcal{H}}(T)|\tilde{\psi}(T)\rangle}{\langle \tilde{\psi}(T)|\tilde{\psi}(T)\rangle}. \tag{4}$$

This gives a real number, which is the energy of one of the two bands if the state is exactly on it or a value in between the energies of the two bands if the state is a superposition. We can further define the band index $\langle \mathcal{B}(T) \rangle = 2\langle E(T)\rangle/(E_+(T) - E_-(T))$ to indicate whether the final state is on the lower-energy band ($\langle \mathcal{B}(T)\rangle = -1$), on the higher-energy band ($\langle \mathcal{B}(T)\rangle = 1$), or a mixture of the two bands ($\langle \mathcal{B}(T)\rangle \approx 0$).

In the Hermitian case, following the black path in Fig. 2 $(a)$, we can change $\tilde{q}_x$ either in the negative direction $\tilde{q}_x: 1 \to -1$ or the positive direction $\tilde{q}_x: -1 \to 1$, where the positions of the initial and final states are close to the band gap at $\tilde{q}_x = 0$. With the initial state loaded on the lower-energy band, the corresponding final band index $\langle \mathcal{B}(T) \rangle$ in the negative and positive directions with respect to $|v|$ is depicted by the black lines in Fig. 2 $(b)$ and $(c)$, respectively. In the limit of a small $|v|$, when we slowly change $\tilde{q}_x$ in either direction, the state can adiabatically follow the band and finally stay on the lower-energy band, $\langle \mathcal{B}(T) \rangle = -1$, which is consistent with the adiabatic theorem [30]. In the other limit of $|v|$, i.e., while rapidly changing $\tilde{q}_x$, the state has insufficient time to change its spin polarization or the relative phase between spin-up and spin-down atoms. As a result, the state ultimately ends in a linear combination of the two bands, $\langle \mathcal{B}(T) \rangle \approx 0$, to reproduce the initial probability distribution instead of completely transitioning to the higher-energy band. This is an example in which a complete interband transition cannot be achieved because the positions of the initial and final states are close to the band gap at $\tilde{q}_x = 0$ [43].

To achieve a complete interband transition when the positions of the initial and final states are close to the band gap, in the non-Hermitian case, the non-Hermitian parameter $\Delta \tilde{\gamma}$ is turned on and off as an interim process to assist interband transitions between two energy bands. In Fig. 2 $(a)$, the green paths can go in either a counterclockwise (CCW) direction $\tilde{q}_x: 1 \to -1$ or a clockwise (CW) direction $\tilde{q}_x: -1 \to 1$. Specifically, for instance, in the CCW direction, we can first increase $\Delta \tilde{\gamma}$ from 0 to 1.2, followed by a decrease in $\tilde{q}_x$ from 1 to $-1$, and finally a decrease in $\Delta \tilde{\gamma}$ to 0. With the initial state loaded on the lower-energy band, the final band index $\langle \mathcal{B}(T) \rangle$ as a function of $|v|$ for the CCW direction and CW direction is depicted by the green lines in Fig. 2 $(b)$ and $(c)$. We note that in the CCW direction, when $|v|$ is relatively small, the state prepared on the lower-energy band can completely transition to the higher-energy band, $\langle \mathcal{B}(T) \rangle = 1$. However, in the CW direction, regardless of the value of $|v|$, the state cannot completely transition to the higher-energy band, $\langle \mathcal{B}(T) \rangle \neq 1$.

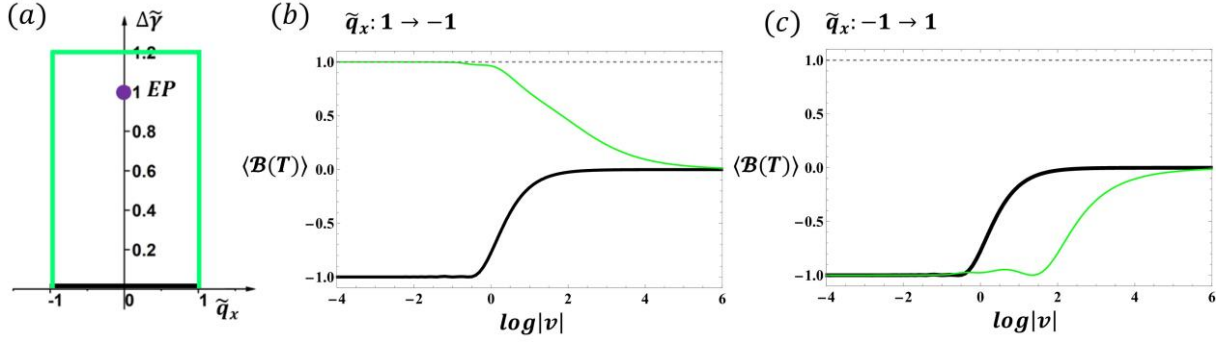

**Fig. 2** Panel $(a)$ shows the paths in the 2D external parameter space $(\tilde{q}_x, \Delta\tilde{\gamma})$. On the black path (Hermitian case), we only tune $\tilde{q}_x$, and on the green path (non-Hermitian case), we tune both $\tilde{q}_x$ and $\Delta\tilde{\gamma}$. The black and green lines in panels $(b)$ and $(c)$ show the final band index $\langle \mathcal{B}(T) \rangle$ if we change the external parameters along the corresponding paths in panel $(a)$ in the CCW and CW directions, respectively. In the Hermitian case, the intraband transitions (adiabatic evolution) occur at small $|v|$, the complete interband transitions cannot occur. In the non-Hermitian case, the complete interband transitions only occur in the CCW direction at small $|v|$.

## 4. Trajectories and Point-Source Diagrams

To clarify the reason for the success of the complete interband transition in the CCW direction and failure in the CW direction, we present detailed numerical results for the full dynamic evolution in the two directions. While the Hamiltonian is non-Hermitian in the interim process, the $\langle E \rangle$ in Eq. (4) has to be extended from a real number to a complex number in order to assess how similar the evolving state is to the two complex-energy bands:

$$\langle E(\tilde{t}) \rangle = \frac{\sum_{i=+,-} c_i^*(\tilde{t}) c_i(\tilde{t}) E_i(\tilde{t})}{\sum_{i=+,-} c_i^*(\tilde{t}) c_i(\tilde{t})}. \tag{5}$$

In this equation, the real and imaginary parts of $\langle E(\tilde{t}) \rangle$ range between the $\mathrm{Re}(E)$ and $\mathrm{Im}(E)$ of the two complex-energy bands, respectively. The complex coefficients for the instantaneous eigenstates $|\tilde{\psi}(\tilde{t})\rangle = c_+(\tilde{t})|\tilde{\psi}_+(\tilde{t})\rangle + c_-(\tilde{t})|\tilde{\psi}_-(\tilde{t})\rangle$ can be extracted using the corresponding left eigenstates $|\tilde{\phi}_\pm\rangle$ (see Appendix I for details), which satisfy a biorthogonal relation with the eigenstates: $\langle \tilde{\phi}_i | \tilde{\psi}_j \rangle = \delta_{i,j}$, $\sum_{i,j} |\tilde{\psi}_i\rangle\langle \tilde{\phi}_i| = I$ ($I$ is an identity matrix) [13,15]. Here, with the norm of $|\tilde{\psi}_\pm\rangle$ equal to 1, we use the symmetry $|\tilde{\psi}_+\rangle = \begin{pmatrix} 0 & 1 \\ -1 & 0 \end{pmatrix} |\tilde{\psi}_-\rangle$ to fix the relative phase between two eigenstates. The band index can still be defined as the same relation in

section 3, i.e., $\langle \mathcal{B}(\tilde{t}) \rangle = 2\langle E(\tilde{t}) \rangle / (E_+(\tilde{t}) - E_-(\tilde{t}))$, which is a real number and can indicate whether the evolving state is on the lower complex-energy band ($\langle \mathcal{B}(\tilde{t}) \rangle = -1$), on the higher complex-energy band ($\langle \mathcal{B}(\tilde{t}) \rangle = 1$), or a mixture of them ($\langle \mathcal{B}(T) \rangle \approx 0$). Here, "higher" and "lower" indicate higher energy and lower energy $(\text{Re}(E))$ in Fig. 1 $(b)$, respectively.

In Fig. 3 $(a)$ and $(b)$, we plot the complex-energy bands, together with the $\text{Re}(\langle E(\tilde{t}) \rangle)$ of the evolving state (green trajectories) when $|v| = e^{-2}$ in the CCW direction and CW direction, respectively. In the CCW direction, we observe that the state adiabatically follows the lower-loss band (colored red, $\langle S \rangle > 0$). When the state crosses the branch cut connecting the lower-loss (red) and higher-loss (blue) bands, it climbs up from the lower complex-energy band to the higher complex-energy band. In contrast, in the CW direction, with the occurrence of a NAT, the state transitions from the higher-loss band to the lower-loss band and then adiabatically follows it. This NAT occurs because the lower-loss band has a smaller damping rate than the higher-loss band. Upon crossing the branch cut, different from the CCW case, the state drops down from the higher complex-energy band to the lower complex-energy band. The asymmetric behavior when crossing the branch cut in different directions explains why the complete interband transition from the lower-energy band to the higher-energy band succeeds only in the CCW direction. We note that a similar behavior can also be observed in references [26, 27].

In addition to the above analysis, we develop a point-source diagram approach that can be used to visualize and predict the occurrence of NATs. The point-source diagram depicts the band index of the evolving state when we change the external parameters from a specific initial state along straight lines in any direction in the 2D external parameter space $(\tilde{q}_x, \Delta \tilde{\gamma})$. These straight lines can be parametrically described by $\tilde{q}_x(\tilde{t}) = \tilde{q}_x(0) + v_q \tilde{t}$ and $\Delta \tilde{\gamma}(\tilde{t}) = \Delta \tilde{\gamma}(0) + v_\gamma \tilde{t}$, with $|v|$ being constant and $\tan^{-1}(v_\gamma / v_q)$ ranging from $0$ to $2\pi$. In Fig. 3 $(c)$, the color of the point-source diagram displays the band index $\langle \mathcal{B}(\tilde{t}) \rangle$ of the evolving states, starting from the lower complex-energy band at $(\tilde{q}_x(0), \Delta \tilde{\gamma}(0)) = (-1,1)$ with $|v| = e^{-2}$. We observe that with keeping away from the initial position $(\tilde{q}_x(0), \Delta \tilde{\gamma}(0))$, the band index of the evolving state $\langle \mathcal{B}(\tilde{t}) \rangle$ gradually changes from $-1$ to $1$ in any direction, which shows that the NATs occur over a process instead of at a time point. Here, we define the locations where the sign of $\langle \mathcal{B}(\tilde{t}) \rangle$ flips as the locations of NATs in the 2D external parameter space.

The black circle in Fig. 3 $(c)$ shows our theoretical model for predicting the locations of NATs. The predicted locations form a circle because we assume that they are solely dependent on the initial position $(\tilde{q}_x(0), \Delta\tilde{\gamma}(0))$, complex coefficients of the initial state $|\tilde{\psi}(0)\rangle = c_+(0)|\tilde{\psi}_+(0)\rangle + c_-(0)|\tilde{\psi}_-(0)\rangle$, and time rate change of the external parameters $|v|$. The predicted radius $R$ of NATs is formulated as follows (see Appendix II for details):

$$R = \frac{|v|}{\text{Im}(\Delta E(0))} \tanh^{-1}\left(\frac{-\xi + \sqrt{\xi^2 - 4\text{Re}(b_i)\text{Re}(b_i n(0)^2)}}{2\text{Re}(b_i n(0)^2)}\right), \quad (6)$$

$$\xi = 1 + |b_i|^2, \quad \Delta E(0) = \frac{E_+(0) - E_-(0)}{2}.$$

In this equation, $b_i \triangleq \frac{c_+(0) - c_-(0)}{c_+(0) + c_-(0)}$, which is equal to $-1$ if the initial state is $|\tilde{\psi}_-(0)\rangle$ and $1$ if the initial state is $|\tilde{\psi}_+(0)\rangle$, becomes a complex number when the initial state comprises both $|\tilde{\psi}_-(0)\rangle$ and $|\tilde{\psi}_+(0)\rangle$; $n(0) = e^{i\tan^{-1}(|v|/2|\Delta E(0)|^3)}$, where the phase factor is associated with the magnitude of the interband Berry connection $|v|/2|\Delta E(0)|^2$ divided by $|\Delta E(0)|$. When $|v|/2|\Delta E(0)|^3 \ll 1$ and therefore $n(0) \approx 1$, the radius $R$ in Eq. (6) is proportional to $|v|$. This suggests that reducing the time rate change of the two external parameters can result in a smaller radius of the NATs. Fig. 3 $(d)$ presents a point-source diagram that shares the same initial state as the case depicted in Fig. 3 $(c)$ but has a velocity $|v| = 0.5e^{-2}$. As expected, the radius of the NATs decreases.

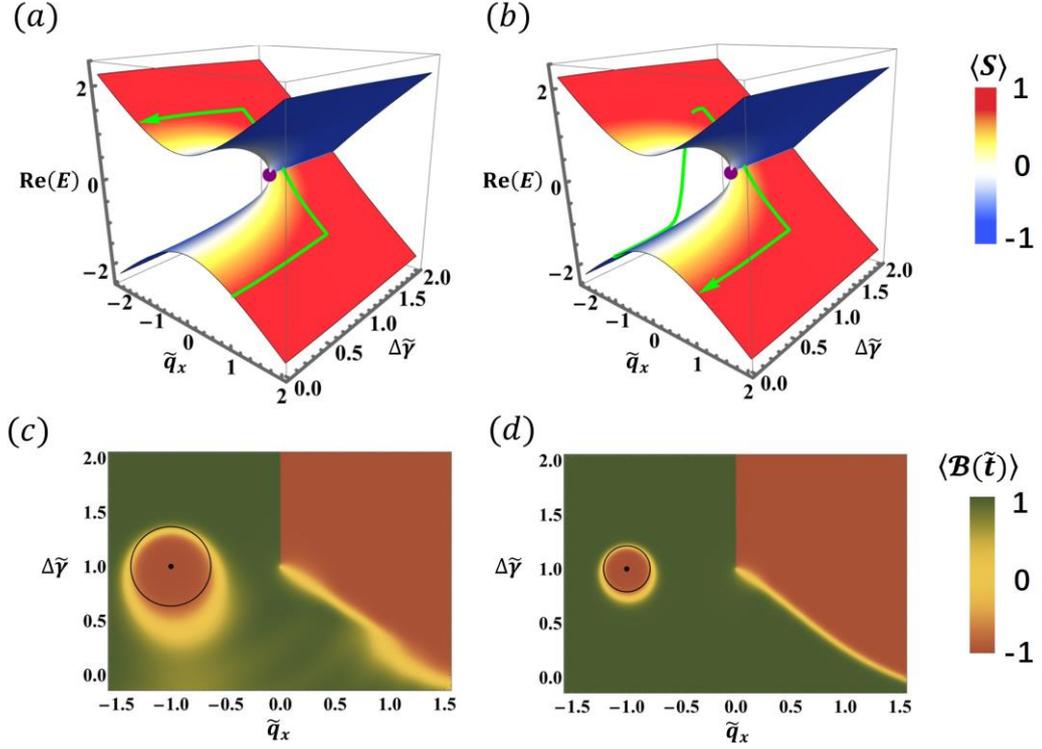

**Fig. 3** The green trajectories $\text{Re}(\langle E(\tilde{t})\rangle)$ in panels $(a)$ and $(b)$ show the dynamic evolution of the atoms in the CCW and CW directions at $|v| = e^{-2}$. The state climbs in the CCW direction and has a NAT and then drops in the CW direction when it crosses the branch cut, which restricts the complete interband transitions in the CCW direction. The point-source diagrams in panels $(c)$ and $(d)$ share the same initial state $|\psi_-(0)\rangle$ at the initial position $(\tilde{q}_x(0), \Delta\tilde{\gamma}(0)) = (-1,1)$ (black center) but have a different velocity, $|v| = e^{-2}$ for panel $(c)$ and $|v| = 0.5e^{-2}$ for panel $(d)$. The black circles show our theoretical model for predicting the locations of NATs.

## 5. Control Protocol of Complete Interband Transitions in Both Directions

As we have discussed, when the external parameters of the system change along the extended non-Hermitian path (green path in Fig. 2 $(a)$), the complete interband transition is restricted to the CCW direction due to the state asymmetric behavior while crossing the branch cut in different directions. In this section, we develop the control protocol to facilitate complete interband transitions in both directions with the green path in Fig. 4 $(a)$. Please notice that we aim to facilitate complete interband transitions in both directions with a single path rather than two independent paths in the 2D external parameter space. The initial and final positions of the state remain in the quasimomentum parameter space (Hermitian Hamiltonian), whereas the non-Hermitian parameter $\Delta\tilde{\gamma}$ will be turned on and off differently from the green path in Fig.

2 (a). For example, in the CCW direction, $\Delta\tilde{\gamma}$ will be increased from 0 to 1.2 and subsequently decreased to 0 after decreasing $\tilde{q}_x$ from 1 to $-1$, rather than being turned on before changing $\tilde{q}_x$. With the initial state prepared on the lower-energy band, the green lines in Fig. 4 (b) and (c) represent the final band index $\langle \mathcal{B}(T) \rangle$ of the state as a function of $|v|$ in the CCW and CW directions. The state completely transitions to the higher-energy band ($\langle \mathcal{B}(T) \rangle = 1$) in both directions when $|v|$ is relatively small. In each panel, the trajectory of $\text{Re}(\langle E(\tilde{t}) \rangle)$ is shown for $|v| = e^{-2}$. The states adiabatically evolve when $\Delta\tilde{\gamma}$ is turned off and transition from the lower complex-energy band to the higher complex-energy band when the non-Hermitian parameter $\Delta\tilde{\gamma}$ is induced. This indicates that the complete interband transitions in both directions are a result of combining adiabatic evolution in the Hermitian system and NATs in the non-Hermitian system. Together with the trivial adiabatic evolution at small $|v|$ [30], it is possible to realize both intraband and complete interband transitions with flexible choice of the positions of the initial and final states in the quasimomentum parameter space.

In addition to varying the time rate change of the two external parameters $|v|$, we also investigate the impact of other factors on the control protocol. Specifically, these factors include the minimum range of the non-Hermitian parameter $\Delta\tilde{\gamma}$ and the position of $\tilde{q}_x$ where $\Delta\tilde{\gamma}$ is induced, which are represented by $h$ and $x_m$ in Fig. 4 (d). Here, in the CCW direction, we first decrease $\tilde{q}_x$ from 1 to $x_m$; then, we increase $\Delta\tilde{\gamma}$ from 0 to $h$ and subsequently decrease it to 0; and finally, we decrease $\tilde{q}_x$ from $x_m$ to $-1$. As the state is assumed to be initially loaded on the lower-energy band, the range of $x_m$ is chosen to be less than 0. This ensures that the higher complex-energy band is the lower-loss band when $\Delta\tilde{\gamma}$ is induced; thus, the target interband transition from the lower-energy band to higher-energy band can occur. Fig. 4 (e) and (f) depict the final band index $\langle \mathcal{B}(T) \rangle$ as a function of $x_m$ and $h$ at $|v| = e^{-3}$ for the CCW and CW directions. It is observed that both figures can be separated into two distinct regions, where the region with $\langle \mathcal{B}(T) \rangle = 1$ indicates the occurrence of complete interband transitions, while the region with $\langle \mathcal{B}(T) \rangle = -1$ denotes the states remaining on the same band. In both directions, complete interband transitions occur with a higher $h$ because a larger $\Delta\tilde{\gamma}$ can provide a larger damping rate contrast for NATs to occur. In addition, a smaller $x_m$ can reduce the $h$ needed to implement the control protocol. The blue lines in Fig. 4 ($e$) and ($f$) show our prediction of the minimum $h$ for the complete interband transitions. We assume that the state remains on the lower complex-energy band at $(x_m, h)$, and we can use the formula in Eq. (6) to predict the radius $R$ of the NATs. To have complete interband transitions, the NAT must occur before $\Delta\tilde{\gamma}$ is reduced to 0, which gives us the condition $R = h$ to plot the blue lines. This shows our point-source diagram approach can facilitate the estimation of the minimum range of loss contrast in the control protocol.

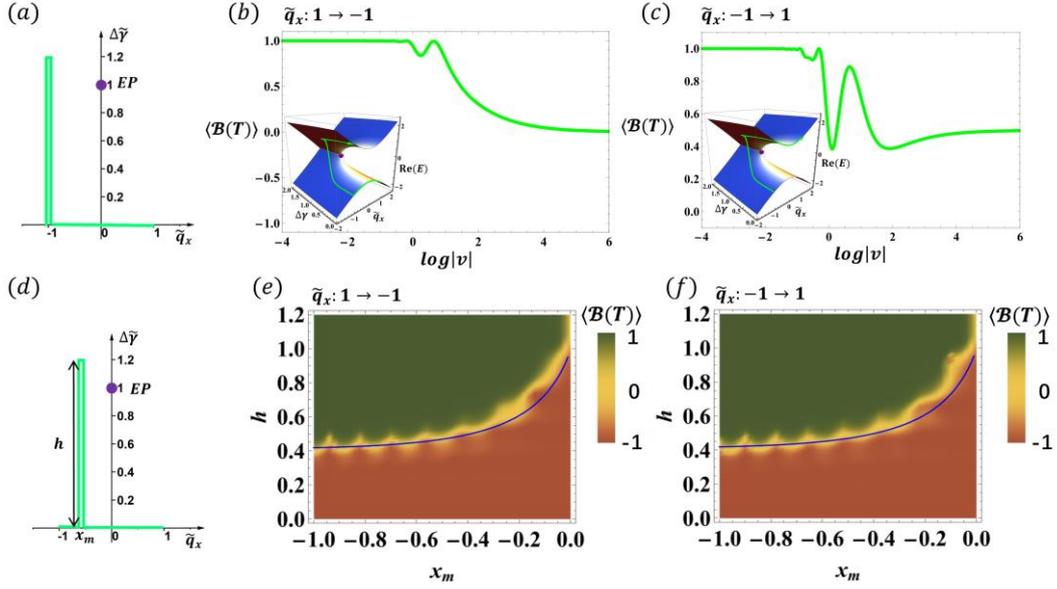

**Fig. 4** Panel ($a$) shows the single path in the 2D external parameter space for realizing the complete interband transitions in both directions. Panels ($b$) and ($c$) show the final band index $\langle\mathcal{B}(T)\rangle$ in the CCW and CW directions. The complete interband transitions ($\langle\mathcal{B}(T)\rangle = 1$) occur in both directions at small $|v|$. Inside each panel, we plot the trajectory $\text{Re}(\langle E(\tilde{t})\rangle)$ at $|v| = e^{-2}$. Panel ($d$) shows the path along which we vary $h$ and $x_m$. Panels ($e$) and ($f$) depict the final band index $\langle\mathcal{B}(T)\rangle$ as a function of $x_m$ and $h$ at $|v| = e^{-3}$ for the CCW and CW directions. In the region $\langle\mathcal{B}(T)\rangle = 1$, the complete interband transitions occur; in the region $\langle\mathcal{B}(T)\rangle = -1$, the states remaining on the same band. The blue line in each panel shows our predictions of the minimum $h$ needed to realize the complete interband transitions using the radius of NATs in Eq. (6).

## 6. Conclusion

In this work, we propose a control protocol for realizing interband transitions by controlling the atom loss in a non-Hermitian spin-orbit-coupled atomic system. Compared with realizing interband transitions by rapidly changing the two-photon detuning, this process does not rely on the assumption that the initial and final states are far from the band gap. Together with the trivial adiabatic evolution, it is possible to realize both intraband and complete interband transitions with flexible choice of the positions of the initial and final states. Our approach can be immediately useful for quantum state control and quantum simulations in a wider range of situations. In addition, the point-source diagram approach introduced in this work enables visualization and prediction of the locations of NATs, facilitating estimation of the minimum range of loss contrast in the control protocol.

## Appendix I. Right and left eigenstates of $\widetilde{\mathcal{H}}$

With choosing $\Omega_R = 2E_r$ in the main text, the Hamiltonian in Eq. (3) can be written as

$$\widetilde{\mathcal{H}} = \begin{pmatrix} -\tilde{q}_x + i\Delta\tilde{\gamma} & 1 \\ 1 & \tilde{q}_x - i\Delta\tilde{\gamma} \end{pmatrix}.$$

The two eigenvalues are $E_\pm = \pm \Delta E = \pm\sqrt{1 + (-\tilde{q}_x + i\Delta\tilde{\gamma})^2}$, with the two right eigenstates (in the main text, we call them eigenstates for simplicity) as

$$|\tilde{\psi}_+\rangle = [\sin(\theta), \cos(\theta)]^T,$$

$$|\tilde{\psi}_-\rangle = [-\cos(\theta), \sin(\theta)]^T,$$

where $\theta = \tan^{-1}(\Delta E - \tilde{q}_x + i\Delta\tilde{\gamma})$. Here we use the square bracket to enclose quantities as a row vector while the transpose operation $T$ turns it into a column vector. When the non-Hermitian parameter $\Delta\tilde{\gamma} = 0$, $\theta$ is a real number and two right eigenstates satisfy the orthogonality and completeness relations $\langle\tilde{\psi}_i|\tilde{\psi}_j\rangle = \delta_{i,j}$, $\sum_{i,j}|\tilde{\psi}_i\rangle\langle\tilde{\psi}_i| = I$ ($I$ is an identity matrix). However, when the non-Hermitian parameter $\Delta\tilde{\gamma} \neq 0$, $\theta$ is a complex number, we have to use the left eigenstates (eigenvectors of $\widetilde{\mathcal{H}}^\dagger$) to fix the orthogonality and completeness. The left eigenstates are

$$|\tilde{\phi}_+\rangle = [\sin^*(\theta), \cos^*(\theta)]^T,$$

$$|\tilde{\phi}_-\rangle = [-\cos^*(\theta), \sin^*(\theta)]^T,$$

which satisfy a biorthogonal relation with the right eigenstates: $\langle\tilde{\phi}_i|\tilde{\psi}_j\rangle = \delta_{i,j}$, $\sum_{i,j}|\tilde{\psi}_i\rangle\langle\tilde{\phi}_i| = I$ [13,15]. Here, $\langle\tilde{\phi}_i|\tilde{\psi}_j\rangle$ means inner product and $|\tilde{\psi}_i\rangle\langle\tilde{\phi}_i|$ means outer product. Thus, with the left eigenstates, we can extract the coefficients of right eigenstates when the Hamiltonian is non-Hermitian.

## Appendix II. Derivation of the locations of NATs

In a time-varying system, after we transform the Hamiltonian ($\Omega_R = 2E_r$) in Eq. (2) from the diabatic basis to the adiabatic basis, it becomes $\widetilde{\mathcal{H}} = \begin{pmatrix} -\Delta E(\tilde{t}) & i\frac{\vartheta(\tilde{t})}{2\,\Delta E(\tilde{t})^2} \\ -i\frac{\vartheta(\tilde{t})}{2\,\Delta E(\tilde{t})^2} & \Delta E(\tilde{t}) \end{pmatrix}$, where $\vartheta(\tilde{t}) = -v_q + iv_\gamma$ and $\Delta E(\tilde{t}) = (E_+(\tilde{t}) - E_-(\tilde{t}))/2$. In the case that the elements in the Hamiltonian can be considered constant under constant $\vartheta$ and $\Delta E(\tilde{t})$ is nearly the same as the initial $\Delta E(0)$, the evolution of $b(\tilde{t}) \triangleq \frac{c_+(\tilde{t}) - c_-(\tilde{t})}{c_+(\tilde{t}) + c_-(\tilde{t})}$ for the state $\psi(\tilde{t}) = c_-(\tilde{t})|\psi_-(\tilde{t})\rangle + c_+(\tilde{t})|\psi_+(\tilde{t})\rangle$ obeys

$$b(\tilde{t}) = \frac{b(0)\frac{\Delta E'(0)}{\Delta E(0)} - i\left(1 - i\frac{\vartheta}{2\Delta E(0)^3}\right)\tan(\Delta E'(0)\tilde{t})}{\frac{\Delta E'(0)}{\Delta E(0)} - ib(0)\left(1 + i\frac{\vartheta}{2\Delta E(0)^3}\right)\tan(\Delta E'(0)\tilde{t})}, \tag{A1}$$

where $\Delta E'(0) = \Delta E(0)\frac{i\sqrt{-\vartheta^2 - 4\Delta E(0)^6}}{2\Delta E(0)^3}$. Assuming $|\vartheta| \ll |\Delta E(0)|$, we only keep the terms to order $\frac{\vartheta}{2\,\Delta E(0)^3}$ (the interband Berry connection divided by $\Delta E(0)$) in the following derivation. Thus, Eq. (A1) can be rewritten as

$$b(\tilde{t}) = \frac{b(0) - i\tan(\Delta E(0)\,\tilde{t})/n(0)}{1 - ib(0)\tan(\Delta E(0)\tilde{t})\,n(0)}, \tag{A2}$$

where $n(0) = e^{i\tan^{-1}(\frac{\vartheta}{2\Delta E(0)^3})}$. As the NATs in the non-Hermitian case are caused mainly by $\mathrm{Im}(\Delta E(0))$ and are nearly independent of the direction of $\vartheta$, we use $|v|$ to replace $\vartheta$ and further neglect $\mathrm{Re}(\Delta E(0))$. Then, $n(0) = e^{i\tan^{-1}(\frac{|v|}{2|\Delta E(0)|^3})}$ and Eq. (A2) becomes

$$b(\tilde{t}) = \frac{b(0) + \tanh(\mathrm{Im}(\Delta E(0))\,\tilde{t})/n(0)}{1 + b(0)\tanh(\mathrm{Im}(\Delta E(0))\tilde{t})\,n(0)}. \tag{A3}$$

For a state that is initially one of the eigenstates, $b(0)$ is either $-1$ or $1$ and $b(\tilde{t})$ is generally a complex number. We can obtain the time point at which NATs occur by solving $\mathrm{Re}(b(\tilde{t}_{occur})) = 0$ (selecting the solution $\tilde{t}_{occur} > 0$) and use the time point and velocity $|v|$ to determine the locations of the NATs, which are given by Eq. (6).